\documentclass[conference,a4paper]{APSIPA2025}
\usepackage{amsmath}
\usepackage{graphicx}
\usepackage{multirow}

\usepackage[utf8]{inputenc}
\usepackage[T1]{fontenc}
\usepackage{microtype}
\usepackage{times}
\usepackage{latexsym}
\usepackage{inconsolata}
\usepackage{amsmath}
\usepackage{amssymb}
\usepackage[table]{xcolor}   
\usepackage{colortbl}        
\usepackage{pgf}             
\usepackage{booktabs}        
\usepackage{multirow}
\usepackage{adjustbox}
\usepackage{verbatim}
\usepackage{subcaption}
\usepackage{algorithm2e}
\usepackage{url}
\usepackage{float}
\usepackage{etoolbox}
\usepackage{tcolorbox}

\definecolor{lightred}{rgb}{1.0, 0.8, 0.8}
\definecolor{lightblue}{rgb}{0.8, 0.9, 1.0}
\definecolor{lightgreen}{rgb}{0.8, 1.0, 0.8}
\definecolor{lightyellow}{rgb}{1.0, 1.0, 0.8}
\definecolor{lightpurple}{rgb}{0.9, 0.8, 1.0}
\definecolor{lightorange}{rgb}{1.0, 0.9, 0.8}

\usepackage{tikz}
\usepackage{xcolor}

\definecolor{mattegreen5}{RGB}{240,248,240}  
\definecolor{mattegreen10}{RGB}{225,242,225} 
\definecolor{mattegreen15}{RGB}{210,235,210} 
\definecolor{mattegreen20}{RGB}{195,228,195} 
\definecolor{mattegreen25}{RGB}{180,220,180} 
\definecolor{mattegreen30}{RGB}{165,213,165} 
\definecolor{mattegreen35}{RGB}{150,205,150} 
\definecolor{mattegreen40}{RGB}{135,198,135} 
\definecolor{mattegreen45}{RGB}{120,190,120} 
\definecolor{mattegreen50}{RGB}{105,183,105} 
\definecolor{mattegreen55}{RGB}{90,175,90}   
\definecolor{mattegreen60}{RGB}{75,168,75}   
\definecolor{mattegreen65}{RGB}{60,160,60}   

\usepackage{threeparttable}
\usepackage[backend=biber,style=ieee,]{biblatex}
\usepackage{geometry}
\usepackage{fancyhdr}

\fancypagestyle{firststyle}{
  \fancyhf{}
  \fancyhead[C]{2025 Asia Pacific Signal and Information Processing Association Annual Summit and Conference (APSIPA ASC)}
}

\begin{document}

\title{Rethinking Cross-Corpus Speech Emotion Recognition Benchmarking: Are Paralinguistic Pre-Trained Representations Sufficient?}

\author{
\authorblockN{
Orchid Chetia Phukan\authorrefmark{2}\authorrefmark{1},
Mohd Mujtaba Akhtar\authorrefmark{2}\authorrefmark{3}\authorrefmark{1},
Girish\authorrefmark{2}\authorrefmark{4}\authorrefmark{1},
Swarup Ranjan Behera\authorrefmark{5}\authorrefmark{1} \\
Parabattina Bhagath\authorrefmark{6},
Pailla Balakrishna Reddy\authorrefmark{7},
Arun Balaji Buduru\authorrefmark{2}
}

\authorblockA{
\authorrefmark{2}IIIT-Delhi, India,
\authorrefmark{3}V.B.S.P.U, India,
\authorrefmark{4}UPES, India,
\authorrefmark{5}Independent Researcher, India \\
\authorrefmark{6}L.B.R College of Engineering, India,
\authorrefmark{7}Reliance AI, India \\
E-mail: \textcolor{blue}{orchidp@iiitd.ac.in}
}
}

\maketitle
\begingroup
  \renewcommand{\thefootnote}{\fnsymbol{footnote}}
  \setcounter{footnote}{0}
  \footnotetext{* Contributed equally as first author.}
\endgroup
\thispagestyle{firststyle}
\pagestyle{fancy}

\begin{abstract}
Recent benchmarks evaluating pre-trained models (PTMs) for cross-corpus speech emotion recognition (SER) have overlooked PTM pre-trained for paralinguistic speech processing (PSP), raising concerns about their reliability, since SER is inherently a paralinguistic task. We hypothesize that PSP-focused PTM will perform better in cross-corpus SER settings. To test this, we analyze state-of-the-art PTMs representations including paralinguistic, monolingual, multilingual, and speaker recognition. Our results confirm that TRILLsson (a paralinguistic PTM) outperforms others, reinforcing the need to consider PSP-focused PTMs in cross-corpus SER benchmarks. This study enhances benchmark trustworthiness and guides PTMs evaluations for reliable cross-corpus SER.
\end{abstract}

\section{Introduction}
Emotions significantly influence human behavior and interactions, and Speech Emotion Recognition (SER) identifies these cues by analyzing speech features such as pitch, tone, and intensity. This capability is valuable across many fields: it enables empathetic responses in human-computer interaction, supports mental health monitoring, enhances customer service through personalized interactions, and adapts learning environments in education. Additionally, in entertainment and security, SER improves user experience and safety by identifying emotional distress or threats. Thus, accurately recognizing emotions in speech is key for building intelligent, responsive systems. \par
\noindent Early research in SER predominantly relied on handcrafted acoustic features, such as MFCCs, paired with traditional machine learning models \cite{milton2013svm, dahake2016speaker, jin2015speech}. With the advent of speech pre-trained models (PTMs), SER has seen significant advancements, as these models provide powerful, generalizable representations learned from large-scale, diverse datasets \cite{atmaja2022evaluating, sharma2022multi, chen2023exploring}. This shift has alleviated both performance benefit and need for training models from scratch. PTMs, however, differ in various aspects, including their training data, which spans across diverse data distributions, and whether they are trained on monolingual or multilingual datasets. Additionally, these models vary in terms of architecture and the pre-training strategies used, such as self-supervised or supervised learning approaches. These differences in PTM nature and pre-training methods have direct implications on their downstream performance for SER. As a result, the choice of PTM can significantly affect the accuracy and robustness of SER and this variability in performance underscores the need for a deeper understanding of these PTMs. \par
\noindent Several benchmarks have been proposed such as SUPERB \cite{yang21c_interspeech}, EMO-SUPERB \cite{wu2024emo}, OPEN-EMOTION \cite{wu2024open} and so on to better understand the performance of different PTMs for SER in monolingual or multilingual settings. These benchmarks also act as a reference for future research for selection of PTMs depending on their use case. However, these benchmarks evaluates PTMs for training and testing on the same corpus. As such there is a recent ongoing interest in the research community to access the cross-corpus SER capability of various SOTA speech PTMs \cite{naini2024generalization, ibrahim24_interspeech}. Here, ``cross-corpus'' encompasses two key scenarios: (1) the same language with varying data distributions and (2) cross-lingual settings. Cross-corpus SER presents greater challenges compared to same-corpus SER due to the domain shift that occurs between the training and test data. 
In such scenarios, the models must generalize across different data distributions, speaker demographics, recording environments, or even languages, which can lead to significant performance degradation. The variations in emotional expression, speaking styles, and acoustic conditions across corpora further complicate the task, making cross-corpus SER more demanding.  \par
\noindent In response, various benchmarks have been proposed in recent couple of years such as SER-evals \cite{osman24_interspeech} and EmoBox \cite{ma24b_interspeech}. However, these benchmarks haven't considered representations from PTM pre-trained primarily for paralinguistic speech processing (PSP). Such oversight raises concerns about the trustworthiness of the benchmarks as comprehensive references for future research, especially since SER fundamentally is a PSP task. Also, Phukan et al. \cite{phukan24b_interspeech} which has shown the topmost performance of paralinguistic PTM representations for SER in multiple languages, haven't evaluated paralinguistic PTM representations for cross-corpus SER. So, to solve this research gap and also to get better understanding, we explore paralinguistic PTM representations for cross-corpus SER. \textit{We hypothesize that representations from paralinguistic PTM representations are better suited for cross-corpus SER. Unlike general-purpose PTM representations, paralinguistic PTM representations captures speech characteristics cues such as intonation, pitch, rhythm, and prosody—features that are directly relevant for SER. These paralinguistic representations transcend linguistic boundaries and can generalize more effectively across different languages and data distributions.} To validate our hypothesis, we perform a comprehensive comparative study of representations from SOTA PTMs comprising paralinguistic, monolingual, multilingual as well as speaker recognition. These PTMs are SOTA in their respective benchmarks. For example, Whisper reported the topmost performance for cross-corpus SER in SER-evals \cite{osman24_interspeech} and EmoBox \cite{ma24b_interspeech}. So, we are presenting the comparison of best of the best PTMs in our study to validate the performance and show the capability of paralinguistic PTM for cross-corpus SER. \newline
\noindent\textbf{Key contributions of the paper are as follows:}

\begin{itemize}
    \item We present a comprehensive comparative study of representations from SOTA PTMs including paralinguistic (TRILLsson), monolingual (WavLM, Unispeech-SAT, Wav2vec2), multilingual (XLS-R, Whisper, MMS) and speaker recognition (x-vector) for cross-corpus SER. We experiment with CREMA-D (\textit{English}), RAVDESS (\textit{English}), emo-DB (\textit{English}), MESD (\textit{Mexican Spanish}), and AESDD (\textit{Greek}) benchmark SER datasets.
    \item Our findings demonstrate TRILLsson (paralinguistic PTM) representations superior performance in cross-corpus SER across all the datasets.
    \item Our study will act as benchmark for future researchers for selection of PTM representations for cross-corpus SER. Also, our study calls for the inclusion of paralinguistic PTM representations in the previous benchmarks for cross-corpus SER. This will ensure reliability and trustworthiness of cross-corpus SER benchmarks, ensuring that future work can build upon more accurate and generalizable results.
\end{itemize}


\begin{figure}[hbt!]
    \centering
    \begin{minipage}[b]{0.25\textwidth}
        \centering
        \includegraphics[width=\textwidth]{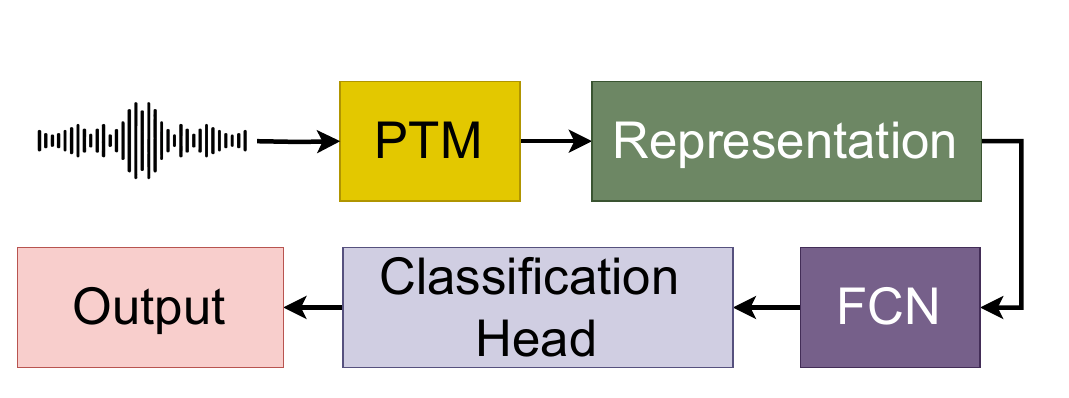}
        \subcaption{Fully Connected Network (FCN)}
        \label{fig:fcn}
    \end{minipage}
    
    \vspace{0.4cm} 
    
    \begin{minipage}[b]{0.35\textwidth}
        \centering
        \includegraphics[width=\textwidth]{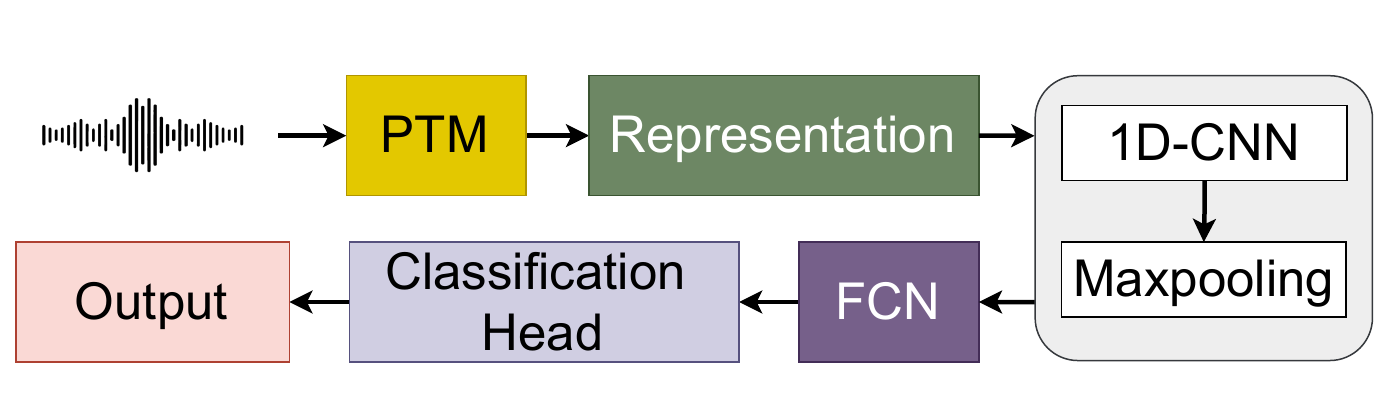}
        \subcaption{Convolutional Neural Network (CNN)}
        \label{fig:cnn}
    \end{minipage}
    \caption{Modeling}
    \label{fig:archi}
\end{figure}

\section{Pre-Trained Models}\label{sec:PTME}
In this section, we discuss the PTMs considered in our study. As monolingual PTMs, we use \textbf{WavLM}\footnote{\url{https://huggingface.co/microsoft/wavlm-base}} \cite{chen2022wavlm}, \textbf{Unispeech-SAT}\footnote{\url{https://huggingface.co/microsoft/unispeech-sat-base}} \cite{chen2022unispeech}, and \textbf{Wav2vec2}\footnote{\url{https://huggingface.co/facebook/wav2vec2-base}} \cite{baevski2020wav2vec}. WavLM is a SOTA PTM in SUPERB and shows top performance compared to other PTMs in various speech processing tasks and trained to perform speech denoising and masked modeling together. Unispeech-SAT also shows SOTA performance in SUPERB and was trained in speaker-aware format. Wav2vec2 is not SOTA like WavLM and Unispeech-SAT in SUPERB, however, we consider it, as it shows relatably good performance in SER as shown by previous research \cite{pepino21_interspeech}. We use the base versions for WavLM, Unispeech-SAT, Wav2vec2 trained on librispeech 960 hours data with 94.70M, 94.70M, and 95.04M parameters respectively. As multilingual PTMs, we consider, \textbf{XLS-R}\footnote{\url{https://huggingface.co/facebook/wav2vec2-xls-r-300m}} \cite{babu22_interspeech}, \textbf{Whisper}\footnote{\url{https://huggingface.co/openai/whisper-base}} \cite{radford2023robust}, and \textbf{MMS}\footnote{\url{https://huggingface.co/facebook/mms-1b}} \cite{pratap2024scaling}. XLS-R was trained on 128 languages while Whisper was trained on 96 languages. XLS-R is based on Wav2vec2 architecture and trained in a self-supervised fashion and in contrast, Whisper is based on vanilla transformer encoder-decoder architecture. We use XLS-R 300M parameters version and for whisper,we use the base variant with 74M. MMS is built on top of Wav2vec2 architecture and improves over Whisper in multilingual speech processing applications. It extends its pre-training to almost over 1400 languages. We use the 1B variant of MMS in our experiments. For paralinguistic PTM, we consider \textbf{TRILLsson} \cite{shor22_interspeech} and it is derived from SOTA paralinguistic Conformer (CAP12) through knowledge distillation. 
TRILLsson is open-sourced, however, CAP12 is not. It has demonstrated SOTA performance in different paralingusitic tasks such as SER, speaker identification, and so on in NOSS benchmark. It was trained on the speech samples of Audioset and Libri-light datasets during its distillation phase. We use the 63.4 million parameters variant of TRILLsson. We also consider \textbf{x-vector}\footnote{\url{https://huggingface.co/speechbrain/spkrec-xvect-voxceleb}} \cite{8461375}, a time delay-neural network trained for speaker recognition. It is trained on the combination of Voxceleb1 + Voxceleb2. x-vector has shown its potential for SER \cite{chetiaphukan23_interspeech}, so we included it in our study. \par

\noindent For each frozen PTM, we extract the last hidden states through average pooling. We get representations of 768 from WavLM, Unispeech-SAT, wav2vec2, 1024 from TRILLsson, 1280 from XLS-R, MMS. For whisper, we discard the decoder and extract representations from its encoder with size 512 same as x-vector. We resample the audios to 16 KHz before passing it to the PTMs.

\begin{table*}[hbt!]
\centering
\begin{adjustbox}{width=\textwidth}
\begin{tabular}{l|cc|cc|cc|cc|cc}
\toprule
\multirow{2}{*}{\textbf{P}} 
& \multicolumn{2}{c|}{\textbf{C(Train) - C(Test)}} 
& \multicolumn{2}{c|}{\textbf{M(Train) - M(Test)}} 
& \multicolumn{2}{c|}{\textbf{E(Train) - E(Test)}} 
& \multicolumn{2}{c|}{\textbf{A(Train) - A(Test)}} 
& \multicolumn{2}{c}{\textbf{R(Train) - R(Test)}} \\ 
\cmidrule(lr){2-11}
& \textbf{FCN} & \textbf{CNN}
& \textbf{FCN} & \textbf{CNN}
& \textbf{FCN} & \textbf{CNN}
& \textbf{FCN} & \textbf{CNN}
& \textbf{FCN} & \textbf{CNN} \\ 
\midrule
MM  
& \cellcolor{mattegreen20}76.00 | 75.82 & \cellcolor{mattegreen20}76.48 | 76.37  
& \cellcolor{mattegreen25}76.39 | 76.19 & \cellcolor{mattegreen35}87.50 | 87.55  
& \cellcolor{mattegreen35}86.95 | 83.45 & \cellcolor{mattegreen35}93.33 | 92.76  
& \cellcolor{mattegreen10}24.15 | 14.58  & \cellcolor{mattegreen15}25.62 | 20.54  
& \cellcolor{mattegreen25}61.98 | 60.84 & \cellcolor{mattegreen30}77.08 | 76.79 \\

Wh 
& \cellcolor{mattegreen15}71.05 | 70.79 & \cellcolor{mattegreen15}71.13 | 70.80  
& \cellcolor{mattegreen10}63.89 | 64.40 & \cellcolor{mattegreen20}68.06 | 68.14
& \cellcolor{mattegreen30}86.67 | 86.69  & \cellcolor{mattegreen35}89.33 | 88.74
& \cellcolor{mattegreen20}53.22 | 48.33 & \cellcolor{mattegreen25}54.55 |  49.62
& \cellcolor{mattegreen30}75.52 | 75.69 & \cellcolor{mattegreen35}79.69 | 79.71 \\

W  
& \cellcolor{mattegreen15}70.02 | 68.97 & \cellcolor{mattegreen20}71.36 | 71.21
& \cellcolor{mattegreen5}50.00 | 50.12 & \cellcolor{mattegreen5}51.39 | 51.15  
& \cellcolor{mattegreen35}89.33 | 86.68  & \cellcolor{mattegreen35}94.67 | 94.88 
& \cellcolor{mattegreen30}81.65 | 75.16 & \cellcolor{mattegreen30}82.64 | 82.26  
& \cellcolor{mattegreen25}72.92 | 73.08 & \cellcolor{mattegreen35}80.73 | 80.72 \\

X  
& \cellcolor{mattegreen20}73.09 | 72.99 & \cellcolor{mattegreen25}74.67 | 74.57
& \cellcolor{mattegreen25}74.31 | 74.43 & \cellcolor{mattegreen30}77.78 | 77.30
& \cellcolor{mattegreen25}78.67 | 75.66 & \cellcolor{mattegreen35}92.00 | 92.37 
& \cellcolor{mattegreen15}39.74 | 35.21 & \cellcolor{mattegreen20}44.63 | 41.80
& \cellcolor{mattegreen25}64.58 | 63.67 & \cellcolor{mattegreen35}82.81 | 82.64 \\

W2 
& \cellcolor{mattegreen10}67.19 | 67.11  & \cellcolor{mattegreen15}69.16 | 68.86
& \cellcolor{mattegreen15}62.50 | 62.50 & \cellcolor{mattegreen20}63.19 | 63.57  
& \cellcolor{mattegreen30}86.67 | 87.57 & \cellcolor{mattegreen35}96.00 | 95.00
& \cellcolor{mattegreen25}68.35 | 61.25 & \cellcolor{mattegreen25}70.25 | 68.64  
& \cellcolor{mattegreen25}71.88 | 71.93 & \cellcolor{mattegreen30}71.88 | 72.09 \\

U  
& \cellcolor{mattegreen10}69.08 | 68.74 & \cellcolor{mattegreen15}69.24 | 69.00  
& \cellcolor{mattegreen5}36.11 | 35.17 & \cellcolor{mattegreen10}40.97 | 40.51  
& \cellcolor{mattegreen20}76.00 | 77.71  & \cellcolor{mattegreen25}80.00 | 70.60 
& \cellcolor{mattegreen20}53.72 | 52.48 & \cellcolor{mattegreen25}64.46 | 64.21  
& \cellcolor{mattegreen25}68.75 | 68.69 & \cellcolor{mattegreen30}76.56 | 76.47 \\

\textbf{T}  
& \cellcolor{mattegreen30}\textbf{79.26} | \cellcolor{mattegreen30}\textbf{79.23} & \cellcolor{mattegreen30}\textbf{79.70} | \cellcolor{mattegreen30}\textbf{79.63}  
& \cellcolor{mattegreen35}\textbf{85.42} | \cellcolor{mattegreen35}\textbf{85.30} & \cellcolor{mattegreen35}\textbf{88.89} | \cellcolor{mattegreen35}\textbf{88.85}  
& \cellcolor{mattegreen35}\textbf{97.56} | \cellcolor{mattegreen35}\textbf{94.65} & \cellcolor{mattegreen35}\textbf{98.67} | \cellcolor{mattegreen35}\textbf{98.13}  
& \cellcolor{mattegreen35}\textbf{91.74} | \cellcolor{mattegreen35}\textbf{91.57} & \cellcolor{mattegreen35}\textbf{93.39} | \cellcolor{mattegreen35}\textbf{93.35}  
& \cellcolor{mattegreen35}\textbf{95.33} | \cellcolor{mattegreen35}\textbf{95.31} & \cellcolor{mattegreen35}\textbf{96.36} | \cellcolor{mattegreen35}\textbf{96.35}  \\

XV 
& \cellcolor{mattegreen10}67.66 | 67.44 & \cellcolor{mattegreen15}68.06 | 68.04  
& \cellcolor{mattegreen25}77.08 | 77.00 & \cellcolor{mattegreen30}81.94 | 81.92  
& \cellcolor{mattegreen35}91.00 | 90.67 & \cellcolor{mattegreen35}92.36 | 92.00 
& \cellcolor{mattegreen30}77.54 | 71.65 & \cellcolor{mattegreen30}81.82 | 80.20  
& \cellcolor{mattegreen30}81.25 | 81.19 & \cellcolor{mattegreen35}87.50 | 87.48 \\
\bottomrule

\multirow{2}{*}{\textbf{}} 
& \multicolumn{2}{c|}{\textbf{C(Train) - M(Test)}} 
& \multicolumn{2}{c|}{\textbf{M(Train) - C(Test)}} 
& \multicolumn{2}{c|}{\textbf{E(Train) - M(Test)}} 
& \multicolumn{2}{c|}{\textbf{A(Train) - M(Test)}} 
& \multicolumn{2}{c}{\textbf{R(Train) - M(Test)}} \\ 
\cmidrule(lr){2-11}
& \textbf{FCN} & \textbf{CNN}
& \textbf{FCN} & \textbf{CNN}
& \textbf{FCN} & \textbf{CNN}
& \textbf{FCN} & \textbf{CNN}
& \textbf{FCN} & \textbf{CNN} \\ 
\midrule
MM  
& \cellcolor{mattegreen10}30.56 | 25.54 & \cellcolor{mattegreen15}32.64 | 28.51  
& \cellcolor{mattegreen5}19.43 | 9.33 & \cellcolor{mattegreen5}20.77 | 10.48    
& \cellcolor{mattegreen10}30.56 | 25.06 & \cellcolor{mattegreen15}31.94 | 29.57  
& \cellcolor{mattegreen5}17.89 | 6.51  & \cellcolor{mattegreen5}19.44 | 13.56
& \cellcolor{mattegreen5}22.92 | 11.22 & \cellcolor{mattegreen10}29.17 | 16.49 \\

Wh  
& \cellcolor{mattegreen15}40.28 | 38.94 & \cellcolor{mattegreen20}42.36 | 41.08  
& \cellcolor{mattegreen5}18.93 | 12.94 & \cellcolor{mattegreen10}21.24 | 13.54  
& \cellcolor{mattegreen10}25.69 | 20.69 & \cellcolor{mattegreen20}38.61 | 32.53  
& \cellcolor{mattegreen15}31.25 | 22.47 & \cellcolor{mattegreen15}31.94 | 24.67  
& \cellcolor{mattegreen10}24.31 | 13.46 & \cellcolor{mattegreen15}28.47 | 17.39 \\

W   
& \cellcolor{mattegreen10}25.69 | 19.79 & \cellcolor{mattegreen15}30.56 | 24.90  
& \cellcolor{mattegreen5}20.40 | 9.55  & \cellcolor{mattegreen10}26.59 | 21.13 
& \cellcolor{mattegreen15}28.47 | 25.98 & \cellcolor{mattegreen20}34.72 | 33.24  
& \cellcolor{mattegreen20}29.17 | 26.93 & \cellcolor{mattegreen25}36.81 | 35.88  
& \cellcolor{mattegreen15}30.56 | 23.33 & \cellcolor{mattegreen20}31.25 | 29.94 \\

X   
& \cellcolor{mattegreen20}35.42 | 31.89 & \cellcolor{mattegreen25}40.28 | 39.68  
& \cellcolor{mattegreen15}27.14 | 19.51 & \cellcolor{mattegreen20}30.45 | 24.66  
& \cellcolor{mattegreen10}21.53 | 16.71 & \cellcolor{mattegreen15}28.47 | 28.86  
& \cellcolor{mattegreen5}18.74 | 10.26 & \cellcolor{mattegreen5}20.83 | 12.69  
& \cellcolor{mattegreen5}22.22 | 12.54 & \cellcolor{mattegreen10}23.61 | 16.29 \\

W2  
& \cellcolor{mattegreen10}27.78 | 23.38 & \cellcolor{mattegreen15}29.17 | 27.65  
& \cellcolor{mattegreen5}13.38 | 12.18 & \cellcolor{mattegreen5}17.55 | 13.00  
& \cellcolor{mattegreen10}27.72 | 22.42 & \cellcolor{mattegreen15}31.25 | 30.97  
& \cellcolor{mattegreen10}25.00 | 20.22 & \cellcolor{mattegreen15}27.78 | 23.57  
& \cellcolor{mattegreen15}29.17 | 27.10 & \cellcolor{mattegreen20}30.56 | 27.40 \\

U   
& \cellcolor{mattegreen10}25.69 | 20.63 & \cellcolor{mattegreen10}25.69 | 20.63  
& \cellcolor{mattegreen5}21.09 | 10.75 & \cellcolor{mattegreen5}22.42 | 12.95  
& \cellcolor{mattegreen10}22.22 | 19.32 & \cellcolor{mattegreen15}27.78 | 26.93 
& \cellcolor{mattegreen5}15.28 | 15.15 & \cellcolor{mattegreen25}34.72 | 34.63  
& \cellcolor{mattegreen5}22.92 | 14.34 & \cellcolor{mattegreen10}22.92 | 16.25 \\

\textbf{T}     
& \cellcolor{mattegreen30}\textbf{45.58} | \cellcolor{mattegreen30}\textbf{44.12} & \cellcolor{mattegreen35}\textbf{51.67} | \cellcolor{mattegreen35}\textbf{47.55}  
& \cellcolor{mattegreen30}\textbf{44.78} | \cellcolor{mattegreen25}\textbf{37.54} & \cellcolor{mattegreen35}\textbf{49.66} | \cellcolor{mattegreen30}\textbf{41.74}  
& \cellcolor{mattegreen35}\textbf{50.14} | \cellcolor{mattegreen35}\textbf{44.92} & \cellcolor{mattegreen35}\textbf{55.56} | \cellcolor{mattegreen30}\textbf{46.56}  
& \cellcolor{mattegreen35}\textbf{49.86} | \cellcolor{mattegreen35}\textbf{43.19} & \cellcolor{mattegreen35}\textbf{53.33} | \cellcolor{mattegreen35}\textbf{49.72} 
& \cellcolor{mattegreen35}\textbf{49.34} | \cellcolor{mattegreen35}\textbf{42.84} & \cellcolor{mattegreen35}\textbf{54.03} | \cellcolor{mattegreen35}\textbf{49.67} \\

XV  
& \cellcolor{mattegreen5}22.22 | 21.55 & \cellcolor{mattegreen15}30.56 | 30.19  
& \cellcolor{mattegreen20}32.10 | 27.75 & \cellcolor{mattegreen25}35.09 | 31.52  
& \cellcolor{mattegreen15}27.78 | 25.31 & \cellcolor{mattegreen20}29.17 | 27.89  
& \cellcolor{mattegreen15}25.14 | 22.37 & \cellcolor{mattegreen20}27.08 | 24.88  
& \cellcolor{mattegreen25}34.72 | 34.17 & \cellcolor{mattegreen30}38.89 | 38.12 \\
\bottomrule
\multirow{2}{*}{\textbf{}} 
& \multicolumn{2}{c|}{\textbf{C(Train) - E(Test)}} 
& \multicolumn{2}{c|}{\textbf{M(Train) - E(Test)}} 
& \multicolumn{2}{c|}{\textbf{E(Train) - C(Test)}} 
& \multicolumn{2}{c|}{\textbf{A(Train) - E(Test)}} 
& \multicolumn{2}{c}{\textbf{R(Train) - E(Test)}} \\ 
\cmidrule(lr){2-11}
& \textbf{FCN} & \textbf{CNN}
& \textbf{FCN} & \textbf{CNN}
& \textbf{FCN} & \textbf{CNN}
& \textbf{FCN} & \textbf{CNN}
& \textbf{FCN} & \textbf{CNN} \\ 
\midrule
MM  
& \cellcolor{mattegreen10}33.33 | 10.42 & \cellcolor{mattegreen25}60.00 | 55.17  
& \cellcolor{mattegreen10}25.33 | 17.19 & \cellcolor{mattegreen20}49.33 | 49.33  
& \cellcolor{mattegreen5}22.82 | 11.67 & \cellcolor{mattegreen10}23.84 | 13.67  
& \cellcolor{mattegreen10}29.85 | 19.11 & \cellcolor{mattegreen15}38.67 | 24.69  
& \cellcolor{mattegreen20}45.33 | 44.67 & \cellcolor{mattegreen25}52.32 | 52.00 \\

Wh 
& \cellcolor{mattegreen20}50.67 | 44.29 & \cellcolor{mattegreen25}62.05 | 60.00  
& \cellcolor{mattegreen10}37.33 | 15.76 & \cellcolor{mattegreen10}40.00 | 19.90  
& \cellcolor{mattegreen15}29.11 | 21.17 & \cellcolor{mattegreen20}34.54 | 22.27  
& \cellcolor{mattegreen15}40.00 | 18.64 & \cellcolor{mattegreen35}76.00 | 71.80 
& \cellcolor{mattegreen25}50.67 | 47.97 & \cellcolor{mattegreen30}64.00 | 55.28 \\

W2 
& \cellcolor{mattegreen10}34.71 | 26.40 & \cellcolor{mattegreen20}49.33 | 45.10  
& \cellcolor{mattegreen25}50.67 | 41.19 & \cellcolor{mattegreen25}52.00 | 47.72  
& \cellcolor{mattegreen15}33.52 | 28.31 & \cellcolor{mattegreen20}37.45 | 36.37 
& \cellcolor{mattegreen30}57.33 | 58.58 & \cellcolor{mattegreen35}72.00 | 69.57  
& \cellcolor{mattegreen5}17.33 | 15.84 & \cellcolor{mattegreen5}20.00 | 16.18 \\

X  
& \cellcolor{mattegreen35}72.00 | 68.11 & \cellcolor{mattegreen35}74.67 | 69.71  
& \cellcolor{mattegreen15}38.67 | 26.19 & \cellcolor{mattegreen15}40.00 | 26.20  
& \cellcolor{mattegreen10}25.10 | 15.27 & \cellcolor{mattegreen10}26.07 | 15.27  
& \cellcolor{mattegreen15}38.57 | 25.69 & \cellcolor{mattegreen20}42.78 | 30.97  
& \cellcolor{mattegreen15}35.74 | 38.67 & \cellcolor{mattegreen20}40.00 | 41.04 \\

W2 
& \cellcolor{mattegreen10}36.00 | 26.30 & \cellcolor{mattegreen15}40.00 | 34.17  
& \cellcolor{mattegreen20}49.33 | 35.29 & \cellcolor{mattegreen20}50.67 | 35.69  
& \cellcolor{mattegreen5}18.02 | 9.78 & \cellcolor{mattegreen10}20.06 | 12.41  
& \cellcolor{mattegreen30}61.33 | 59.75 & \cellcolor{mattegreen35}77.33 | 74.43  
& \cellcolor{mattegreen25}52.00 | 40.11 & \cellcolor{mattegreen25}52.00 | 46.80\\

U  
& \cellcolor{mattegreen5}21.33 | 19.52 & \cellcolor{mattegreen10}32.00 | 32.28   
& \cellcolor{mattegreen15}38.67 | 31.32 & \cellcolor{mattegreen20}49.33 | 41.37  
& \cellcolor{mattegreen5}18.80 | 11.56 & \cellcolor{mattegreen10}21.64 | 13.99  
& \cellcolor{mattegreen10}33.33 | 31.02 & \cellcolor{mattegreen15}41.33 | 33.87  
& \cellcolor{mattegreen5}14.07 | 17.33 & \cellcolor{mattegreen10}22.66 | 25.33 \\

\textbf{T}    
& \cellcolor{mattegreen35}\textbf{75.60} | \cellcolor{mattegreen35}\textbf{73.61} & \cellcolor{mattegreen35}\textbf{79.33} | \cellcolor{mattegreen35}\textbf{74.16}  
& \cellcolor{mattegreen25}\textbf{55.33} | \cellcolor{mattegreen20}\textbf{47.03} & \cellcolor{mattegreen30}\textbf{58.00} | \cellcolor{mattegreen30}\textbf{56.73}  
& \cellcolor{mattegreen25}\textbf{48.54} | \cellcolor{mattegreen25}\textbf{46.15} & \cellcolor{mattegreen30}\textbf{54.84} | \cellcolor{mattegreen30}\textbf{52.75}  
& \cellcolor{mattegreen35}\textbf{86.14} | \cellcolor{mattegreen35}\textbf{85.33} & \cellcolor{mattegreen35}\textbf{91.08} | \cellcolor{mattegreen35}\textbf{90.67} 
& \cellcolor{mattegreen35}\textbf{78.67} | \cellcolor{mattegreen35}\textbf{76.38} & \cellcolor{mattegreen35}\textbf{80.00} |  \cellcolor{mattegreen35}\textbf{78.70}\\

XV 
& \cellcolor{mattegreen25}56.00 | 44.84 & \cellcolor{mattegreen30}64.00 | 51.60  
& \cellcolor{mattegreen25}52.00 | 46.90 & \cellcolor{mattegreen30}54.67 | 49.27  
& \cellcolor{mattegreen10}28.09 | 20.44 & \cellcolor{mattegreen15}29.58 | 21.38  
& \cellcolor{mattegreen20}47.25 | 41.25 & \cellcolor{mattegreen25}52.19 | 49.28  
& \cellcolor{mattegreen35}68.80 | 65.99 & \cellcolor{mattegreen35}72.00 | 68.27 \\

\bottomrule

\multirow{2}{*}{\textbf{}} 
& \multicolumn{2}{c|}{\textbf{C(Train) - A(Test)}} 
& \multicolumn{2}{c|}{\textbf{M(Train) - A(Test)}} 
& \multicolumn{2}{c|}{\textbf{E(Train) - A(Test)}} 
& \multicolumn{2}{c|}{\textbf{A(Train) - C(Test)}} 
& \multicolumn{2}{c}{\textbf{R(Train) - A(Test)}} \\ 
\cmidrule(lr){2-11}
& \textbf{FCN} & \textbf{CNN}
& \textbf{FCN} & \textbf{CNN}
& \textbf{FCN} & \textbf{CNN}
& \textbf{FCN} & \textbf{CNN}
& \textbf{FCN} & \textbf{CNN} \\ 
\midrule
MM   
& \cellcolor{mattegreen10}35.54 | 26.33 & \cellcolor{mattegreen15}45.45 | 35.56  
& \cellcolor{mattegreen5}19.83 | 6.62 & \cellcolor{mattegreen5}23.14 | 11.80   
& \cellcolor{mattegreen25}47.11 | 41.94 & \cellcolor{mattegreen35}65.29 | 62.80  
& \cellcolor{mattegreen5}18.65 | 8.60 & \cellcolor{mattegreen10}20.61 | 15.45   
& \cellcolor{mattegreen20}36.36 | 32.85 & \cellcolor{mattegreen25}45.45 | 45.09 \\

Wh  
& \cellcolor{mattegreen15}42.15 | 38.31 & \cellcolor{mattegreen20}44.63 | 44.10  
& \cellcolor{mattegreen5}23.14 | 14.80 & \cellcolor{mattegreen10}31.40 | 22.18  
& \cellcolor{mattegreen25}51.24 | 46.89 & \cellcolor{mattegreen25}52.89 | 46.93  
& \cellcolor{mattegreen10}28.32 | 16.33 & \cellcolor{mattegreen15}34.62 | 26.10  
& \cellcolor{mattegreen15}29.75 | 24.61 & \cellcolor{mattegreen15}31.40 | 25.79 \\

W2  
& \cellcolor{mattegreen5}23.61 | 19.69 & \cellcolor{mattegreen15}45.45 | 40.09  
& \cellcolor{mattegreen15}38.84 | 29.54 & \cellcolor{mattegreen20}40.50 | 32.51  
& \cellcolor{mattegreen20}44.63 | 39.60 & \cellcolor{mattegreen30}55.37 | 50.97  
& \cellcolor{mattegreen15}31.24 | 27.97 & \cellcolor{mattegreen15}32.49 | 28.93  
& \cellcolor{mattegreen20}33.88 | 29.61 & \cellcolor{mattegreen20}39.67 | 33.52 \\

X   
& \cellcolor{mattegreen25}48.76 | 42.90 & \cellcolor{mattegreen30}55.37 | 49.50  
& \cellcolor{mattegreen5}19.18 | 6.62 & \cellcolor{mattegreen10}24.79 | 17.53   
& \cellcolor{mattegreen15}36.36 | 29.04 & \cellcolor{mattegreen25}47.93 | 44.72  
& \cellcolor{mattegreen10}25.85 | 17.40 & \cellcolor{mattegreen10}27.69 | 23.15  
& \cellcolor{mattegreen5}23.97 | 17.46 & \cellcolor{mattegreen15}31.40 | 26.11 \\

W2  
& \cellcolor{mattegreen10}27.27 | 17.64 & \cellcolor{mattegreen15}37.19 | 30.99  
& \cellcolor{mattegreen10}25.62 | 15.54 & \cellcolor{mattegreen20}41.32 | 37.63  
& \cellcolor{mattegreen25}51.24 | 47.91 & \cellcolor{mattegreen25}52.07 | 49.55  
& \cellcolor{mattegreen10}22.90 | 15.28 & \cellcolor{mattegreen10}25.96 | 20.07  
& \cellcolor{mattegreen25}39.67 | 36.58 & \cellcolor{mattegreen25}40.50 | 39.28 \\

U   
& \cellcolor{mattegreen10}29.75 | 24.79 & \cellcolor{mattegreen15}32.23 | 26.49  
& \cellcolor{mattegreen15}28.10 | 24.59 & \cellcolor{mattegreen20}38.02 | 32.76  
& \cellcolor{mattegreen20}38.02 | 32.38 & \cellcolor{mattegreen25}42.98 | 38.41 
& \cellcolor{mattegreen10}25.18| 20.75 & \cellcolor{mattegreen10}25.89  | 23.12  
& \cellcolor{mattegreen5}22.31 | 15.51 & \cellcolor{mattegreen10}28.93 | 26.55 \\

\textbf{T}   
& \cellcolor{mattegreen30}\textbf{62.98} | \cellcolor{mattegreen25}\textbf{55.09} & \cellcolor{mattegreen35}\textbf{67.93} | \cellcolor{mattegreen30}\textbf{61.78} 
& \cellcolor{mattegreen25}\textbf{52.23} | \cellcolor{mattegreen20}\textbf{44.48} & \cellcolor{mattegreen30}\textbf{62.98} | \cellcolor{mattegreen25}\textbf{54.73}  
& \cellcolor{mattegreen30}\textbf{66.12} | \cellcolor{mattegreen30}\textbf{65.36} & \cellcolor{mattegreen35}\textbf{73.90} | \cellcolor{mattegreen35}\textbf{73.55}
& \cellcolor{mattegreen25}\textbf{50.67} | \cellcolor{mattegreen20}\textbf{45.94} & \cellcolor{mattegreen25}\textbf{53.19} | \cellcolor{mattegreen25}\textbf{49.29}  
& \cellcolor{mattegreen25}\textbf{51.40} | \cellcolor{mattegreen20}\textbf{44.32} & \cellcolor{mattegreen30}\textbf{58.84} | \cellcolor{mattegreen25}\textbf{55.14} \\

XV  
& \cellcolor{mattegreen15}37.19 | 29.27 & \cellcolor{mattegreen25}47.11 | 38.54  
& \cellcolor{mattegreen20}38.02 | 33.50 & \cellcolor{mattegreen25}50.41 | 43.70  
& \cellcolor{mattegreen30}61.16 | 60.27 & \cellcolor{mattegreen35}67.77 | 67.23  
& \cellcolor{mattegreen15}31.25 | 26.15 & \cellcolor{mattegreen15}32.65 | 29.65  
& \cellcolor{mattegreen25}46.28 | 38.97 & \cellcolor{mattegreen30}53.72 | 49.69 \\

\bottomrule

\multirow{2}{*}{\textbf{}} 
& \multicolumn{2}{c|}{\textbf{C(Train)-R(Test)}} 
& \multicolumn{2}{c|}{\textbf{M(Train) - R(Test)}} 
& \multicolumn{2}{c|}{\textbf{E(train) - R(Test)}} 
& \multicolumn{2}{c|}{\textbf{A(Train) - R(Test)}} 
& \multicolumn{2}{c}{\textbf{R(train) - C(test)}} \\  
\cmidrule(lr){2-11}
& \textbf{FCN} & \textbf{CNN}
& \textbf{FCN} & \textbf{CNN}
& \textbf{FCN} & \textbf{CNN}
& \textbf{FCN} & \textbf{CNN}
& \textbf{FCN} & \textbf{CNN} \\ 
\midrule
MM   
& \cellcolor{mattegreen10}31.77| 18.93 & \cellcolor{mattegreen15}38.54 | 28.60  
& \cellcolor{mattegreen5}20.31 | 7.67 & \cellcolor{mattegreen10}27.08 | 19.64   
& \cellcolor{mattegreen5}20.31 | 6.75  & \cellcolor{mattegreen25}40.62 | 40.85  
& \cellcolor{mattegreen5}17.65 | 11.90 & \cellcolor{mattegreen10}19.79 | 15.69  
& \cellcolor{mattegreen15}26.28 | 18.75 & \cellcolor{mattegreen20}35.25 | 30.87 \\

Wh  
& \cellcolor{mattegreen35}62.50 | 60.84 & \cellcolor{mattegreen35}64.58 | 61.40  
& \cellcolor{mattegreen5}20.31 | 8.69 & \cellcolor{mattegreen5}21.30 | 8.75  
& \cellcolor{mattegreen10}25.00 | 15.80 & \cellcolor{mattegreen20}37.50 | 34.46  
& \cellcolor{mattegreen10}22.40 | 10.57 & \cellcolor{mattegreen15}29.17 | 27.07 
& \cellcolor{mattegreen20}36.90 | 32.64 & \cellcolor{mattegreen20}38.79 | 36.73 \\

W   
& \cellcolor{mattegreen10}29.17 | 23.17 & \cellcolor{mattegreen15}32.29 | 27.46  
& \cellcolor{mattegreen15}30.73 | 23.06 & \cellcolor{mattegreen20}35.94 | 27.42  
& \cellcolor{mattegreen15}32.29 | 21.84 & \cellcolor{mattegreen15}32.29 | 25.27  
& \cellcolor{mattegreen10}26.56 | 19.30 & \cellcolor{mattegreen15}28.80 | 23.38 
& \cellcolor{mattegreen10}25.89 | 17.06 & \cellcolor{mattegreen15}33.28 | 23.24 \\

X   
& \cellcolor{mattegreen20}40.62 | 28.50 & \cellcolor{mattegreen25}42.19 | 29.18  
& \cellcolor{mattegreen5}20.31 | 7.67 & \cellcolor{mattegreen10}25.52 | 18.09   
& \cellcolor{mattegreen10}22.92 | 11.65 & \cellcolor{mattegreen25}40.62 | 28.50  
& \cellcolor{mattegreen10}24.98 | 18.64 & \cellcolor{mattegreen10}26.04 | 21.66  
& \cellcolor{mattegreen15}31.00 | 24.70 & \cellcolor{mattegreen20}36.66 | 30.48 \\

W2  
& \cellcolor{mattegreen5}21.35 | 8.93 & \cellcolor{mattegreen10}26.56 | 18.47   
& \cellcolor{mattegreen15}29.17 | 21.53 & \cellcolor{mattegreen15}30.73 | 26.00  
& \cellcolor{mattegreen20}34.90 | 27.62 & \cellcolor{mattegreen20}34.90 | 29.65  
& \cellcolor{mattegreen15}29.17 | 27.07 & \cellcolor{mattegreen15}32.29 | 32.24  
& \cellcolor{mattegreen5}19.75 | 10.54 & \cellcolor{mattegreen10}20.61 | 17.97 \\

U   
& \cellcolor{mattegreen20}41.15 | 32.45 & \cellcolor{mattegreen25}43.75 | 37.60  
& \cellcolor{mattegreen5}19.79 | 6.61  & \cellcolor{mattegreen5}19.79 | 6.61   
& \cellcolor{mattegreen5}21.35 | 8.81  & \cellcolor{mattegreen5}21.35 | 8.81  
& \cellcolor{mattegreen15}27.08 | 19.87 & \cellcolor{mattegreen20}31.25 | 26.72  
& \cellcolor{mattegreen15}28.87 | 26.77 & \cellcolor{mattegreen20}32.42 | 31.13 \\

\textbf{T}   
& \cellcolor{mattegreen35}\textbf{71.88} | \cellcolor{mattegreen35}\textbf{71.33} & \cellcolor{mattegreen35}\textbf{72.40} | \cellcolor{mattegreen35}\textbf{72.29}  
& \cellcolor{mattegreen25}\textbf{45.52} | \cellcolor{mattegreen20}\textbf{34.70} & \cellcolor{mattegreen30}\textbf{48.65} | \cellcolor{mattegreen25}\textbf{38.98}  
& \cellcolor{mattegreen30}\textbf{66.15} | \cellcolor{mattegreen30}\textbf{66.12} & \cellcolor{mattegreen35}\textbf{72.40} | \cellcolor{mattegreen35}\textbf{72.29}  
& \cellcolor{mattegreen30}\textbf{64.06} | \cellcolor{mattegreen30}\textbf{63.17} & \cellcolor{mattegreen35}\textbf{69.52} | \cellcolor{mattegreen35}\textbf{69.27}  
& \cellcolor{mattegreen25}\textbf{57.20} | \cellcolor{mattegreen25}\textbf{56.57} & \cellcolor{mattegreen30}\textbf{62.23} | \cellcolor{mattegreen30}\textbf{61.59} \\

XV  
& \cellcolor{mattegreen15}35.52 | 27.30 & \cellcolor{mattegreen20}41.15 | 32.68  
& \cellcolor{mattegreen20}34.38 | 29.53 & \cellcolor{mattegreen25}39.58 | 35.03  
& \cellcolor{mattegreen25}37.84 | 31.91 & \cellcolor{mattegreen25}38.54 | 36.40  
& \cellcolor{mattegreen20}34.89 | 31.45 & \cellcolor{mattegreen20}37.50 | 34.43  
& \cellcolor{mattegreen25}37.84 | 34.80 & \cellcolor{mattegreen25}37.92 | 36.40 \\

\bottomrule
\end{tabular}
\end{adjustbox}
\caption{Evaluation scores of different models trained with different PTM representations for both same corpus and cross-corpus SER; Scores are presented in Accuracy | Macro F1 format; P, C, M, E, A, R stands for PTM, CREMA-D, MESD, Emo-DB, AESDD, and RAVDESS; X(Train) - X(Test) represents the training and evaluation dataset where X = C, M, E, A, R. For example, C(Train) - M(Test) represents training on CREMA-D and testing on MESD; MMS (MM), Whisper (Wh), WavLM (W), XLSR (X), Wav2Vec2 (W2), UniSpeech (U), TRILLsson (T), and X-vector (XV) are PTMs; The intensity highlights the performance levels, where darker shades indicate higher values and lighter shades indicate lower values. The scores are highlighted block-wise i.e. X(Train) - X(Test) (where X = C, M, E, A, R) block based on its relative performance.}
\label{results}
\end{table*}

\section{Experiments}
\subsection{Benchmark Datasets}
\noindent\textbf{Crowd-Sourced Emotional Multimodal Actors Dataset (CREMA-D) \cite{cao2014crema}}: It serves as a widely-used benchmark for SER and it is gender-balanced. It includes recordings in english from 48 male and 43 female actors, totaling 7,442 utterances. The dataset is valuable due to its representation of a variety of speaker ages and ethnic backgrounds. It covers six emotions: angry, happiness, sadness, fear, disgust, and a neutral state, with each actor delivering 12 distinct sentences. \par
\noindent \textbf{Ryerson Audio-Visual Database of Emotional Speech and Song (RAVDESS) \cite{livingstone2018ryerson}}: It contains 7,356 english clips covering eight emotions: neutral, calm, happiness, sadness, angry, fearful, disgust, and surprise for speech, and six emotions for song. The database provides high emotional validity, as validated by 319 raters who assessed recordings on emotional category, intensity, and genuineness. \par
\noindent\textbf{German Emotional Speech Database (Emo-DB) \cite{burkhardt2005database}}: It is recorded in German, contains 535 utterances from five male and five female actors. Each actor was assigned one of ten predefined scripts for the recordings. Emo-DB includes seven emotional categories: angry, fear, boredom, disgust, happiness, neutral, and sadness.\par \noindent\textbf{Mexican Emotional Speech Database (MESD) \cite{duville2021mexican}}: It is a culturally tailored emotional speech dataset designed for Mexican Spanish speakers. It includes 864 utterances featuring six emotional states: angry, disgust, fear, happiness, neutral, and sadness. The recordings are categorized into three demographic groups: male adults, female adults, and children, with non-professional actors delivering carefully selected single-word utterances. \par
\noindent\textbf{Acted Emotional Speech Dynamic Database (AESDD) \cite{vryzas2018speech}}: It is a Greek-language speech emotion dataset consisting of approximately 600 utterances recorded by five actors. It includes five emotional states: angry, disgust, fear, happiness, and sadness. In our study, we consider the common emotions: happiness, fear, sadness, angry, and disgust as we are primarily interested in cross-corpus SER.

\subsection{Downstream Modeling}
We use FCN and CNN as our downstream models as used by previous research in SER \cite{phukan24b_interspeech} and related applications \cite{charola23_interspeech}. As we are interested in understanding the PTMs implicit capability for cross-corpus SER, so we kept the downstream modeling as simple as possible. For CNN (Figure \ref{fig:cnn}), we use 1D-CNN layer with a kernel size of 3 followed by max-pooling with a pool size of 2. The extracted features are then flattened and passed through a FCN block containing a dense layer of 90 neurons. Finally, a classification that contains the output layer with softmax activation function. For FCN (Figure \ref{fig:fcn}), we use the same modeling as used by the FCN block in CNN modeling. The FCN models trainable parameters are between 0.4 to 0.8M and for CNN, it is between 1 to 2M. \par
\noindent \textbf{Training Details}: We use Adam optimizer with batch size of 32 and  cross-entropy as loss function. We train the all the models for 20 epochs. We use dropout and early stopping to prevent overfitting. We follow a five-fold cross-validation for training and evaluation of the models for both same-corpus and cross-corpus experiments. 

\subsection{Results}

Table \ref{results} presents the evaluation results for the downstream models trained with different PTMs. For same-corpus evaluation, we present the average performance across five folds. For cross-corpus evaluation, we follow a similar five-fold approach: in each fold, we train on four folds of the training dataset and test on one fold of the testing dataset. This process is repeated five times, and we report the average performance across these folds to ensure robustness. \par

\begin{figure}[hbt!]
    \centering
    \subfloat[]{%
        \includegraphics[width=0.23\textwidth]{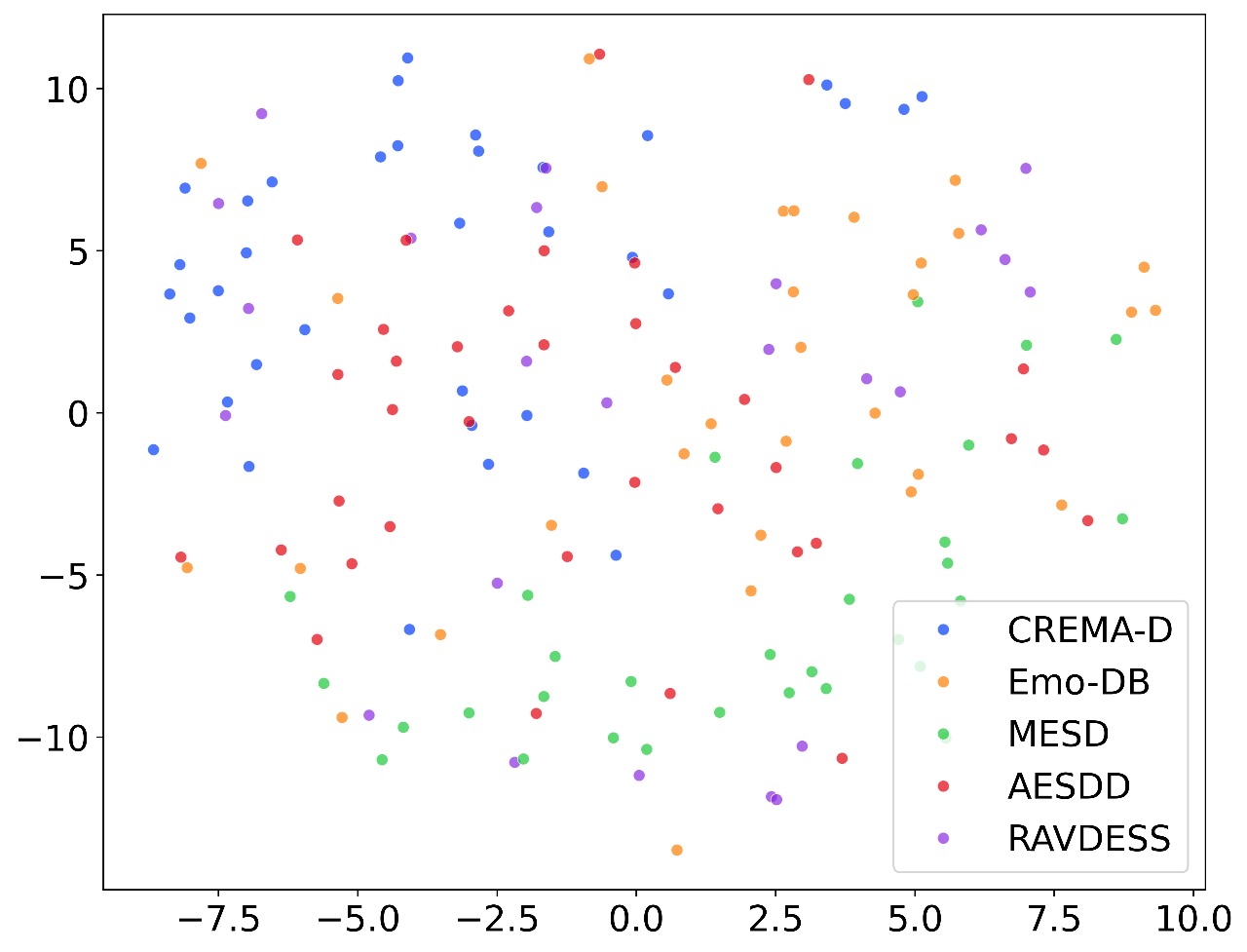}
    }
    \subfloat[]{%
        \includegraphics[width=0.23\textwidth]{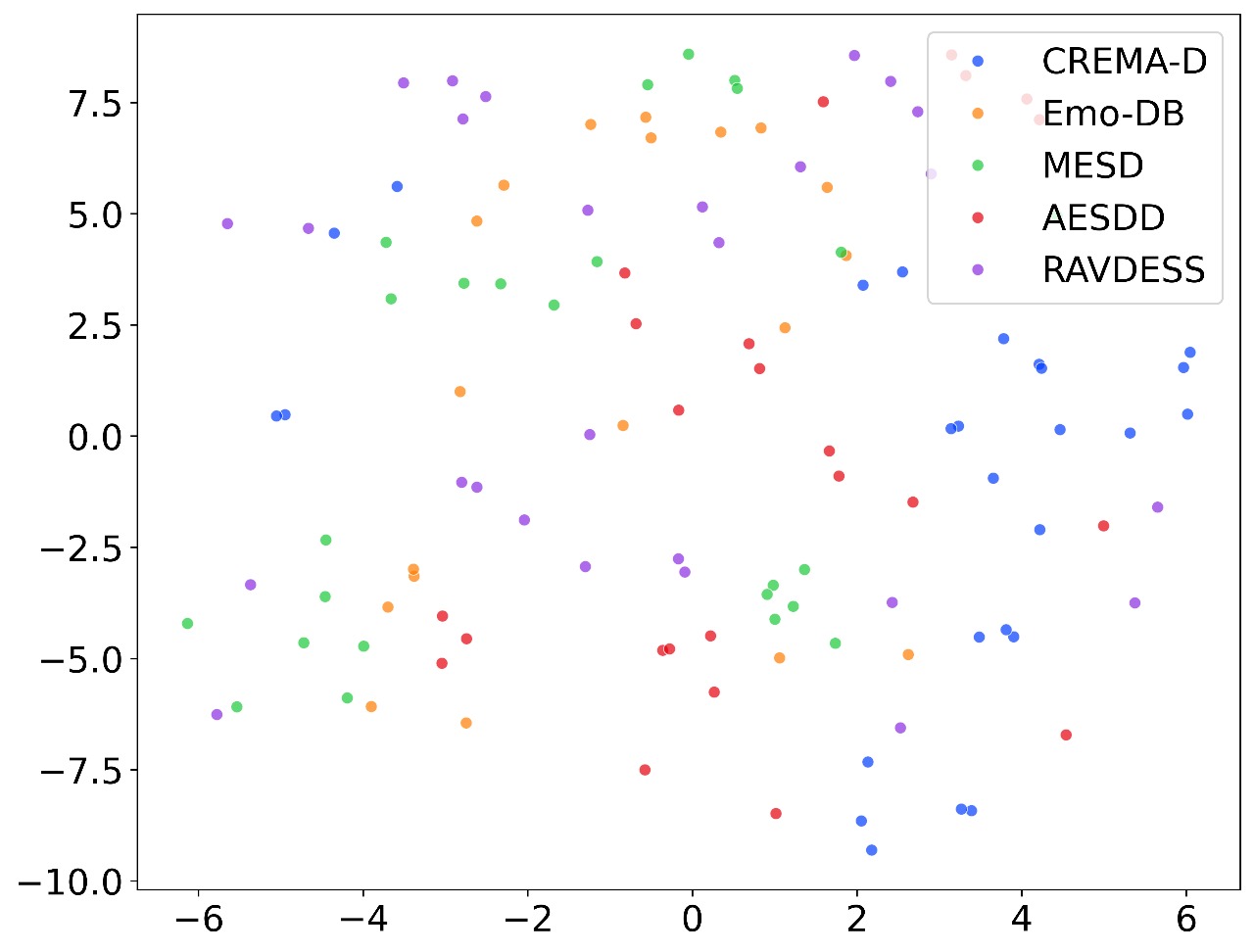}
    }
    \hfill
    \subfloat[]{%
        \includegraphics[width=0.23\textwidth]{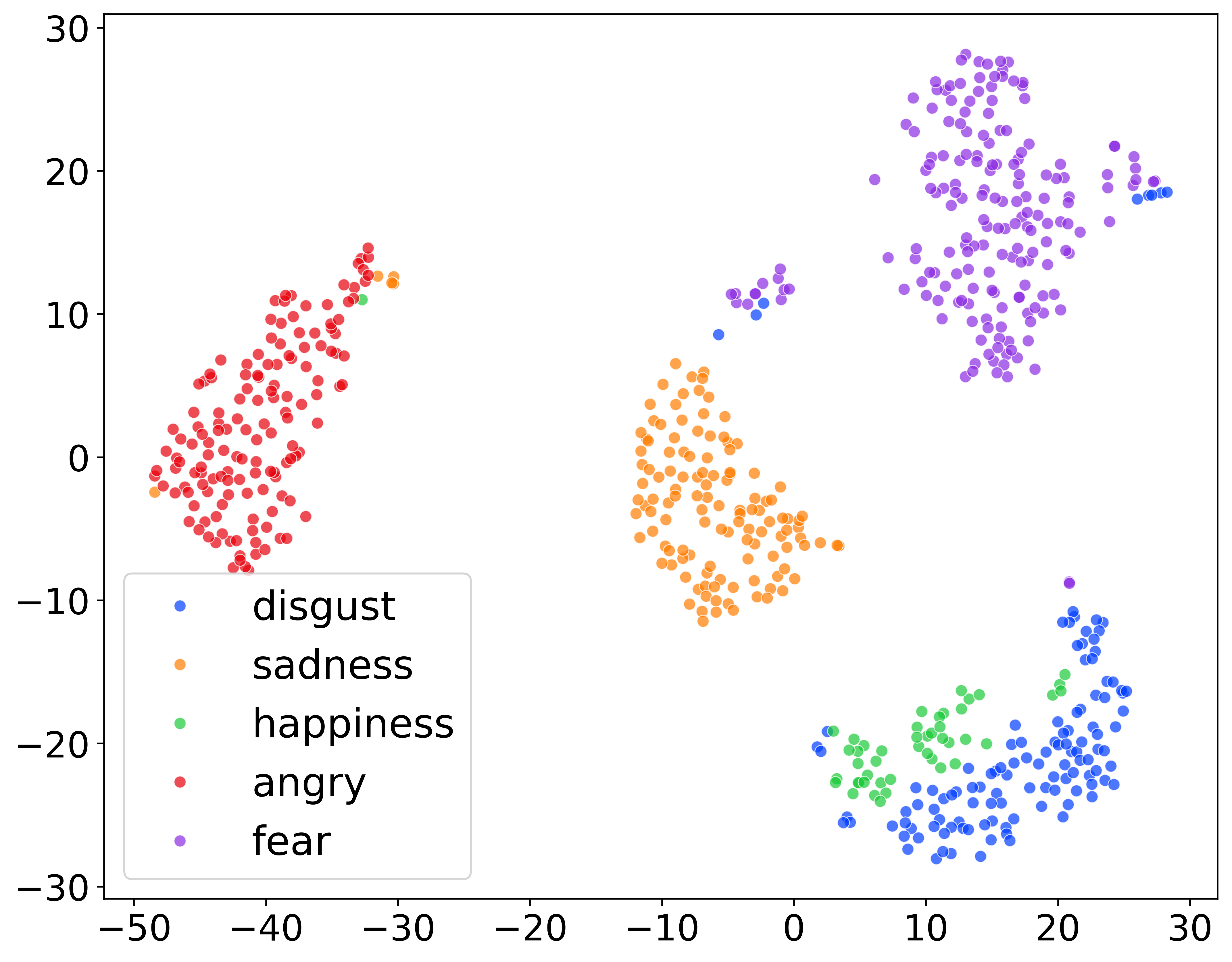}
    }
    \subfloat[]{%
        \includegraphics[width=0.23\textwidth]{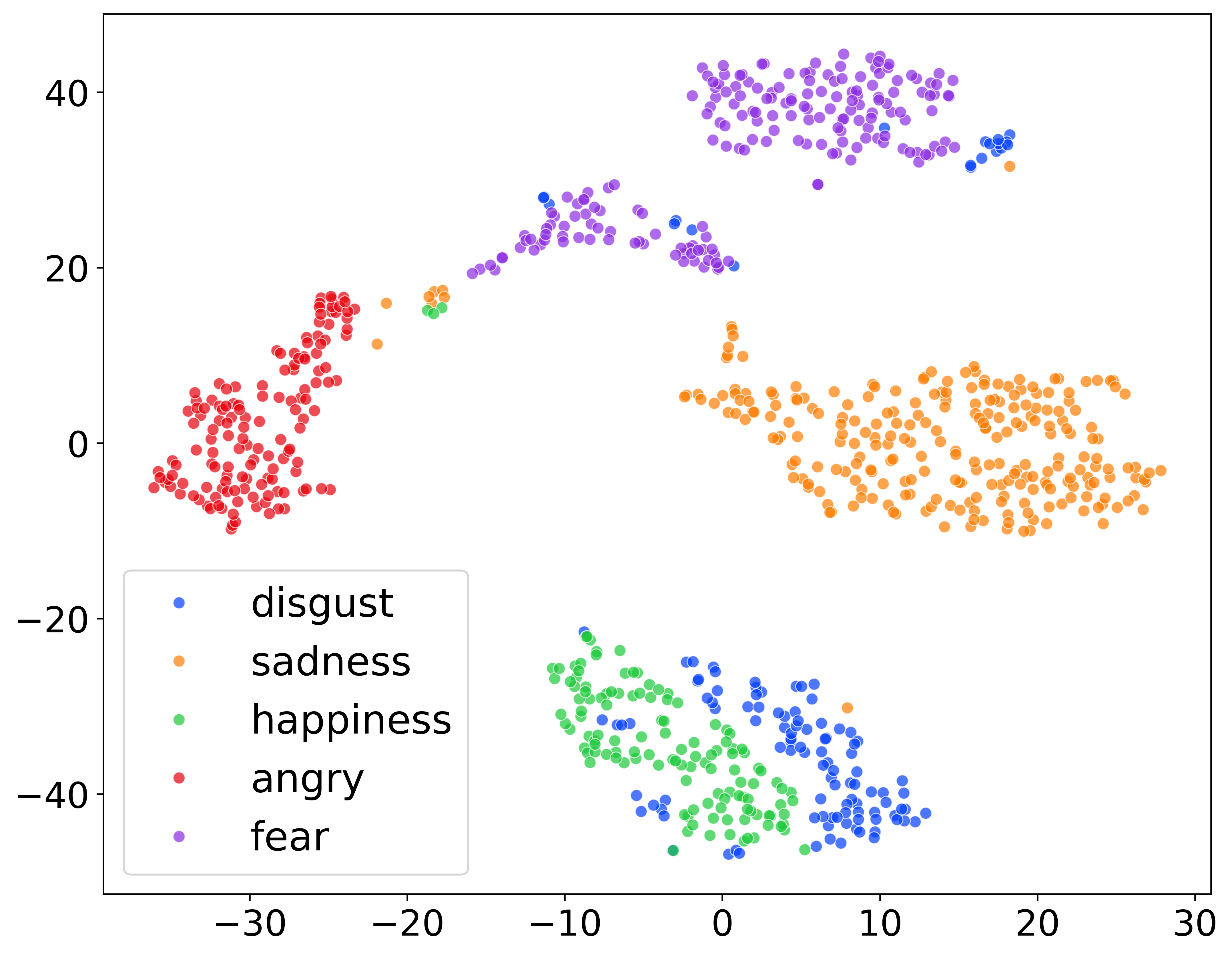}
    }
    \caption{ t-SNE plots of PTMs raw representations: (a) Anger (b) Sadness (c) RAVDESS (d) CREMA-D}
    \label{fig:tsne}
\end{figure}

\begin{figure}[hbt!]
    \centering
    \begin{minipage}[b]{0.20\textwidth}
        \centering
        \includegraphics[width=\textwidth]{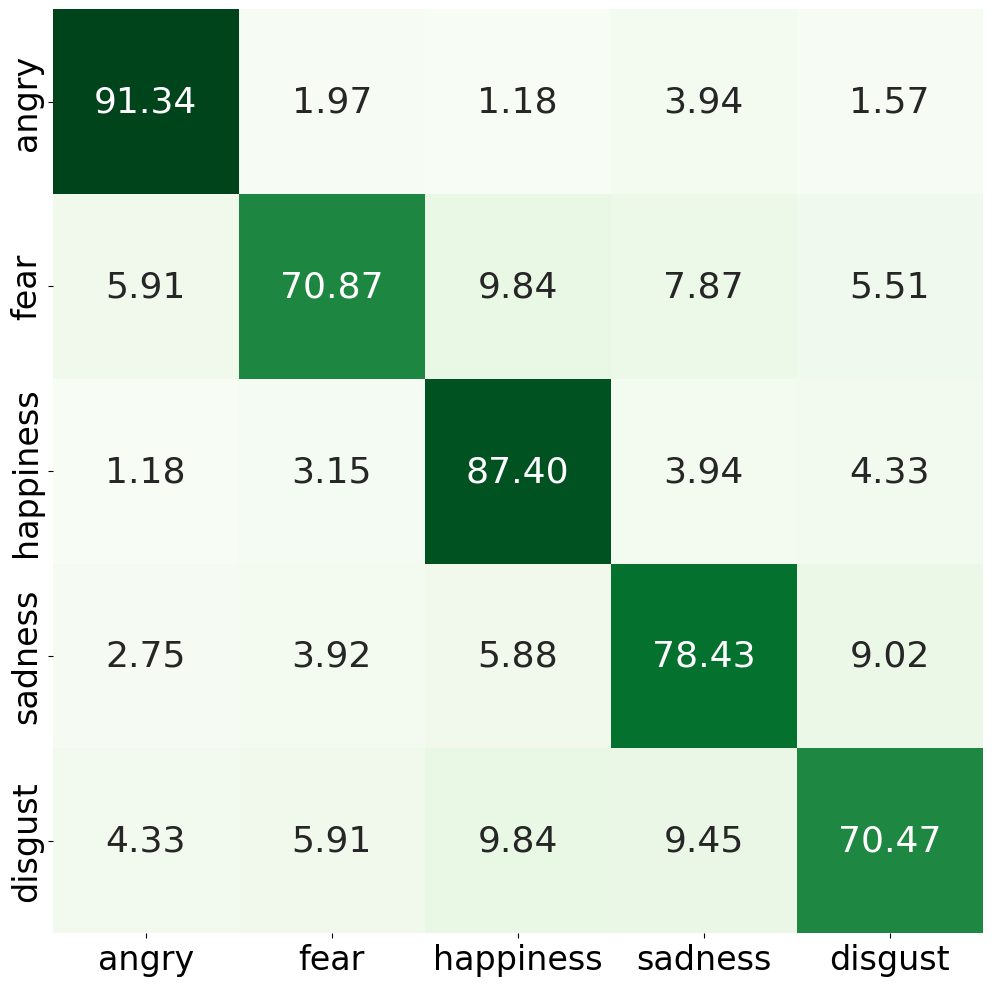}
        \subcaption{}
        \label{fig:fig1}
    \end{minipage}
    \hspace{0.02\textwidth}
    \begin{minipage}[b]{0.20\textwidth}
        \centering
        \includegraphics[width=\textwidth]{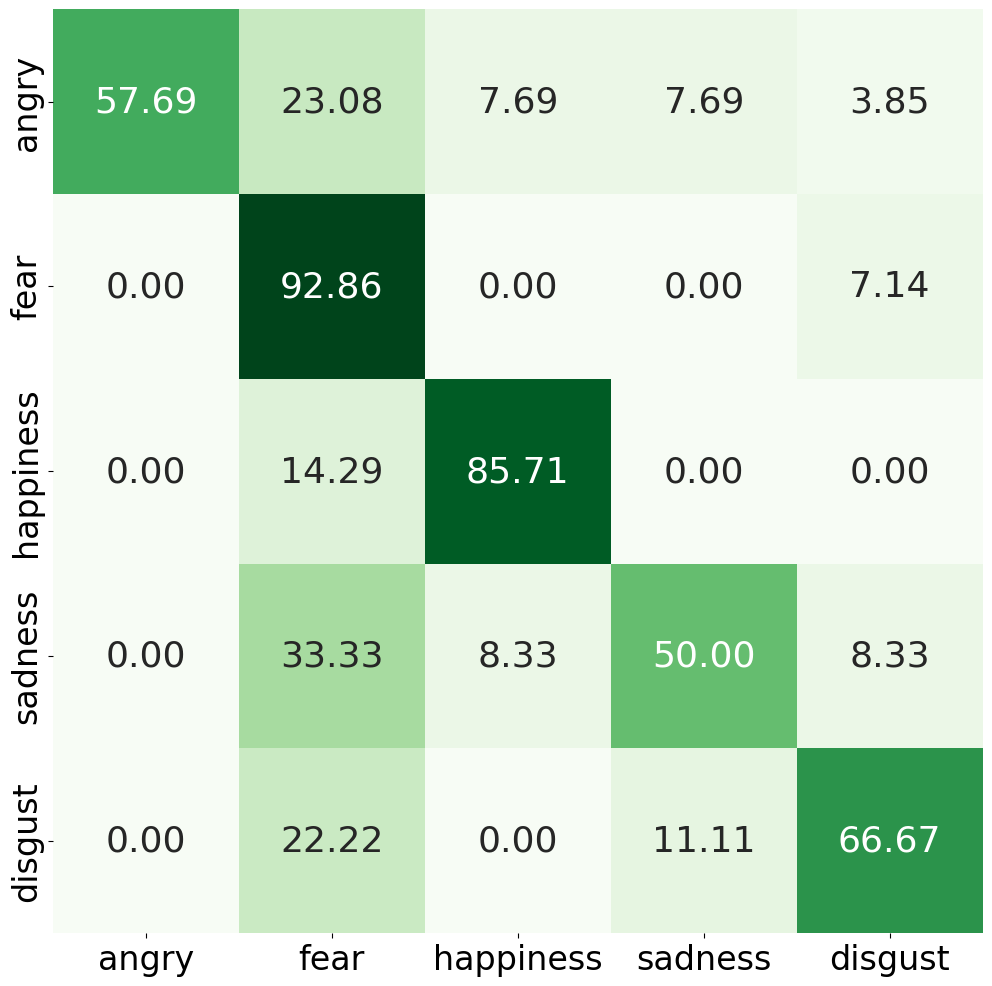}
        \subcaption{}
        \label{fig:fig2}
    \end{minipage}
    \caption{Confusion matrices: (a) Trained and Tested on CREMAD- with TRILLsson representations and CNN downstream (b) Trained on CREMA-D and tested on Emo-DB with TRILLsson representations and CNN downstream; x-axis: Predicted, y-axis: True}
    \label{fig:confusion_matrices}
\end{figure}
\noindent\textbf{For same-corpus experiments} : TRILLsson consistently outperforms monolingual, multilingual, and speaker recognition PTMs across CREMA-D, MESD, Emo-DB, AESDD, and RAVDESS. This aligns with prior studies on evaluating TRILLsson for SER \cite{phukan24b_interspeech}, reinforcing its strength in capturing critical paralinguistic features like pitch, intonation, and rhythm. Among monolingual PTMs, no clear winner emerges, as performance varies across datasets. WavLM outperforms the others in some cases, but Wav2vec2 achieves better results in MESD, indicating sensitivity to data distribution. Similarly, multilingual PTMs exhibit performance variations, with MMS leading for Emo-DB with CNN downstreams, while WavLM (monolingual PTM) surpasses multilingual PTMs in AESDD. Interestingly, x-vector shows mixed performance, occasionally outperforming monolingual and multilingual PTMs but underperforming in other cases, further highlighting the influence of dataset-specific factors. To further validate TRILLsson representations effectiveness, we visualize its raw representations using t-SNE for RAVDESS and CREMA-D in Figure \ref{fig:tsne} (c) and (d), revealing well-separated emotion clusters. Additionally, we present the confusion matrix of CNN downstream trained on TRILLsson representations for CREMA-D for a particular fold in Figure \ref{fig:confusion_matrices} (a), confirming its superior performance.

\noindent\textbf{For cross-corpus experiments} : TRILLsson outperforms other PTMs—including monolingual, multilingual, and speaker recognition PTMs across all datasets. In some instances, it reports top performance by a very large margin in comparison to other PTMs. This validates our \textit{hypothesis that paralinguistic PTM representations are inherently better suited for cross-corpus SER. Unlike other PTMs, TRILLsson focuses on capturing essential speech characteristics—such as intonation, pitch, rhythm, and prosody-that are fundamental for SER in a much better way}. Because these features are largely independent of linguistic content, TRILLsson’s representations exhibit greater robustness across diverse datasets, enabling superior generalization in cross-corpus settings. Among monolingual PTMs, performance varies across datasets. WavLM excels in some cases, while Wav2vec2 achieves better in some, indicating that monolingual PTMs struggle with consistent cross-corpus generalization. Similarly, no single multilingual PTM dominates across all datasets among the multilingual PTMs. These results emphasize that PTM performance in cross-corpus settings is strongly influenced by the underlying data distribution. Despite these multilingual PTMs are pre-trained on multiple languages, they show poor cross-corpus generalization and thus, amplifying the dependence of SER on paralinguistic features and in which TRILLsson excels in. Interestingly, x-vector demonstrates mixed performance, occasionally surpassing both monolingual and multilingual PTMs in some datasets while underperforming in others. This suggests that paralinguistic characteristics embedded in x-vector representations can be beneficial for cross-corpus SER in certain contexts but are not universally effective. We also plot the t-SNE plot visualizations of TRILLsson representations for anger and sadness emotion across all the datasets in Figure \ref{fig:tsne} (a) and (b) respectively. We segmented the anger and sadness emotions for all the datasets, extracted representations from TRILLsson and visualized it using t-SNE plot. We observe that no clear clusters are present and through this, we can say that TRILLsson converts the same emotion samples to a joint representational space irrespective of linguistic difference. These plots further amplifies our obtained results. We also plot the confusion matrix of CNN model trained on CREMA-D with TRILLsson representations and evaluated on Emo-DB in Figure \ref{fig:confusion_matrices} (b).

\vspace{-0.3cm}
\section{Conclusion}
In this study, we investigate paralinguistic PTM (TRILLsson) for cross-corpus SER, which have been largely overlooked by previous benchmarks on evaluating PTMs for cross-corpus SER. We hypothesize that paralinguistic PTM would outperform other PTMs. By analyzing SOTA PTM representations, including paralinguistic, monolingual, multilingual, and speaker recognition, we show that TRILLsson, consistently outperforms other PTMs and validating our hypothesis. These results highlight the importance of our work and it will act as benchmark for future works on cross-corpus SER. It also calls for incorporating paralinguistic PTM in previous cross-corpus SER benchmarks, ultimately enhancing their trustworthiness and offering valuable insights for future PTM evaluations in SER tasks.

\printbibliography

\end{document}